\documentstyle[proceedings,numreferences,epsfig]{crckapb}

\begin{opening}
\title{HERA transverse polarimeter absolute scale and error by rise-time 
calibration}
\author{V.Gharibyan}                                                            
\institute{DESY,~Deutsches~Elektronen~Synchrotron,~Hamburg,~Germany }           
\institute{Yerevan~Physics~Institute,~Yerevan,~Armenia }                        
                                                                                
\author{K.~P.~Sch\"uler}                                                        
\institute{DESY,~Deutsches~Elektronen~Synchrotron,~Hamburg,~Germany }           
\end{opening}
\runningtitle{HERA TRANSVERSE POLARIMETER}
\begin{document}

\begin{abstract}

We give the results of an analysis of some 18 rise-time calibrations
which are based on data collected in 1996/97. Such measurements are used
to determine the absolute polarization scale of the transverse electron 
beam polarimeter (TPOL) at HERA. The results of the 1996/97 calibrations
are found to be in good agreement with earlier calibrations of the TPOL
performed in 1994 with errors of 1.2\% and 1.1\%. Based on these 
calibrations and a comparison with measurements from the longitudinal 
polarimeter (LPOL) at HERA carried out over a two-months period in 2000, 
we obtain a mean LPOL/TPOL ratio of 1.018. Both polarimeters are found
to agree with each other within their overall errors of about 2\% each.

\end{abstract}

\section{Introduction}

Two polarimeters are employed at the HERA ep storage ring to measure the
polarization of its 27.5~GeV electron or positron beam. Both instruments
are laser backscattering Compton devices. The TPOL polarimeter measures
the transverse beam polarization by detecting the associated angular
anisotropy of the backscattered Compton photons. The original 
configuration of this instrument has been covered in considerable 
detail~\cite{tpol_01,tpol_02,tpol_03,Westphal,Schuler} and recent
upgrades are described in~\cite{upgrades}. The LPOL polarimeter measures
the longitudinal beam polarization between the spin rotators at the
HERMES experiment by detecting an asymmetry in the energy spectra of the
Compton photons~\cite{lpol}.

In this paper, we will present the results of an analysis of rise-time 
calibration data which were collected in 1996/97. Such measurements are 
used to determine the absolute polarization scale of the TPOL polarimeter.
These results will be compared with earlier TPOL calibrations obtained in
1994 and with recent cross calibrations with the LPOL polarimeter.

\section{The rise-time calibration method}

Electrons or positrons are injected unpolarized at 12 GeV into the HERA
storage ring and are subsequently ramped to the nominal beam energy of
27.5 GeV. Transverse polarization evolves then naturally through the
spin flip driven by synchrotron radiation
(the Sokolov-Ternov effect~\cite{ST}) with an exponential time dependence

\begin{equation}
P(t) \; = \; P^{\infty} \left( 1 - \exp\left( -t / \tau \right) \right)
\label{eq:pol}
\end{equation}

For a circular machine with a perfectly flat orbit the spin vector of the
positrons (electrons) will be exactly parallel (antiparallel) to the
direction of the guide field and the theoretical maximum of the
polarization has been calculated to be 
$P^{\infty}_{ST} = 8/(5 \sqrt{3}) = 92.4\%$, 
with an associated rise-time constant

\begin{equation}
\tau_{ST}\; = \; P^{\infty}_{ST} \: 
\frac{m_e |\rho|^3}{r_e \hbar \gamma^5}
\label{eq:tau_SK}
\end{equation}

\noindent
where $\gamma$ is the Lorentz factor, $\rho$ is the radius of curvature of
the orbit and the other symbols have the usual meaning.

For rings such as HERA with the spin rotators needed to get
longitudinal polarization at experiments and/or reversed horizontal
bends, $P^{\infty}_{ST}$ can be reduced substantially below $92.4\%$ and
$\tau_{ST}$ can be modified too, see Table 1.  
Synchrotron radiation also causes depolarization which
competes with the Sokolov-Ternov effect with the result that the
equilibrium polarization is reduced even further. Moreover the
depolarization is strongly enhanced by the presence of the small but
non-vanishing misalignments of the magnetic elements and the resulting
vertical orbit distortions which are typically 1~mm rms.

These effects are treated  in the formalism of Derbenev and
Kondratenko \cite{DK} which has been summarized in \cite{Barber}.
Then the equilibrium polarization and the time constant can be written as

\begin{equation}
P^{\infty}\; = \; - \frac{8}{5\sqrt{3}} \, \frac
{{\oint ds \, \langle \, |\rho (s)|^{-3} \, \hat{b} \cdot 
(\hat{n} - \partial \hat{n}/ \partial \delta ) \, \rangle}_s}
{{\oint ds \, \langle \, |\rho (s)|^{-3} \, 
[1 - \frac{2}{9} (\hat{n} \cdot \hat{s})^2 + 
\frac{11}{18} (\partial \hat{n}/ \partial \delta )^2] \, \rangle}_s} 
\label{eq:P}
\end{equation}  

\begin{equation}
\tau \; = \; \frac{8}{5\sqrt{3}} \, \frac {m_e}{r_e \hbar \gamma^5} 
\, C \, \frac {1}
{{\oint ds \, \langle \, |\rho (s)|^{-3} \,
[1 - \frac{2}{9} (\hat{n} \cdot \hat{s})^2 +
\frac{11}{18} (\partial \hat{n}/ \partial \delta )^2] \, \rangle}_s}
\label{eq:tau}
\end{equation}

\noindent
where the unit vector $\hat{n}$ describes the polarization direction
which is a function of the machine azimuth $s$ and the phase space
coordinate $\vec{u} = (x, p_x, y, p_y, z, \delta)$, the unit vectors 
$\hat{b}$ and $\hat{s}$ describe the magnetic field orientation and the
direction of motion, and $C$ is the circumference of the machine.
The angular brackets $\langle \: \rangle_s$ denote an average over
phase space at azimuth $s$. The term with 
$(\partial \hat{n}/\partial \delta)^2 $ accounts for 
the radiative depolarization due to photon-induced longitudinal recoils
and the term with 
$\partial \hat{n}/\partial \delta$ in the numerator of \ref{eq:P}
arises from the dependence of the radiation power on the spin orientation.
 
These expressions can be summarized in the scaling relation
\begin{equation}
\frac{P^{\infty}}{(P^{\infty}_{ST} + \Delta)} \; = \; \frac{\tau}{\tau_{ST}} 
\label{eq:pol-tau}
\end{equation}
between the actually observed parameters $P^{\infty}$ and $\tau$ of 
equation~\ref{eq:pol} and the theoretical values
$P^{\infty}_{ST}$ and $\tau_{ST}$
which are obtained by ignoring terms with 
$\partial \hat{n}/\partial \delta$ in equations \ref{eq:P} and
\ref{eq:tau}. For the HERA machine at 27.5~GeV they take the values in Table 1 
~\cite{private}:

\begin{table}[hbt]
\begin{center}
\begin{tabular}{cccc}\hline
~~HERA status~~&~~~~year~~~~& $\; P^{\infty}_{ST} \;$ &
$ \; \tau_{ST}~(min) \;$ \\
\hline
flat          & 1994    &   0.915               &    36.7          \\
non-flat      & 1996/97 &   0.891               &    36.0          \\
\hline
              &         &                       &                  \\
\end{tabular}
\caption{Input parameters for rise-time calibrations}
\end{center}
\end{table}

\noindent
The correction $\Delta$ in equation \ref{eq:pol-tau} results from the 
term $\partial \hat{n}/\partial \delta$ in the numerator of equation 
\ref{eq:P}.
In the case of a flat ring, $\Delta$ is negligible
compared to $P^{\infty}_{ST}$. However, for a non-flat HERA, namely with
spin rotators activated at HERMES, $\Delta/P^{\infty}_{ST}$ remains small
but can still be significant.  Since the  magnitude of $\Delta$
depends on the distortions of the machine, it is very difficult to
calculate reliably ~\cite{private}.

The scaling relationship of equation \ref{eq:pol-tau} can be exploited to 
predict  the expected equilibrium polarization $P^{\infty}$ when we know the
remaining quantities $P^{\infty}_{ST}$, $\tau_{ST}$, $\tau$ and
$\Delta$.  Through comparison of the actually measured value of
$P^{\infty}$ with the predicted value from equation \ref{eq:pol-tau}
we can therefore calibrate the polarimeter.

This is the essence of the rise-time method. It requires knowledge
of the theoretical maximum values for the machine in the absence of
depolarizing spin diffusion effects, i.e. $P^{\infty}_{ST}$ and
$\tau_{ST}$ from  equations \ref{eq:P}
and \ref{eq:tau}, and a measurement of the actual rise-time $\tau$ from a
fit of data to the functional form of equation \ref{eq:pol}.
For a non-flat ring we also need to know $\Delta$.
However, by comparison with rise-time
calibrations obtained earlier with a flat machine, we will be able to
examine this term experimentally.

\section{Experimental Procedure}

In order to take rise-time calibration data which are free from major
systematic effects it is essential that the build-up of polarization
proceeds under conditions of an extremely stable machine performance
since even minor operator adjustments in the course of a measurement
may change the parameters of the functional form that describes the
time dependence. For this reason rise-time calibrations require 
dedicated HERA operation, where the machine is brought to a very
stable condition and is then only monitored without further operator
invention. The beam polarization is then destroyed by the resonant
depolarization technique by applying an rf field to a weak kicker
magnet~\cite{Zetsche}. When the baseline polarization $P_0$ has been
established the depolarizing rf is turned off at time $t = 0$ and the
subsequent exponential rise is measured under completely quiet machine
conditions. After one or two hours, the polarization can be destroyed
again for another rise-time measurement and so forth. 

In order to retain only data of high quality, we applied the following
selection criteria: (a) stable machine and polarimeter conditions;
(b) depolarizing rf frequency shifts by no more than 100~Hz (corresponding
to a change in beam energy of about 1~MeV); in order to apply this
test the beam needs to be depolarized before and after each rise-time
measurement; (c) the depolarizer should be activated for about 10~minutes
prior to $t = 0$ to establish a reliable baseline $P_0$. Of the rise-time
data collected in 1996/97, altogether 18 curves out of a total of 25
survived these cuts. For the older measurements taken in 1994, 8 curves
out of 14 could be retained~\cite{Barrow}.

\section{Results and Errors}

Figure 1 shows two examples of rise-time data obtained in 1997.
The curves are fits to the following functional form, with
$\tau$, $P_0$ and $K$ as free fitting parameters

\begin{equation}
K \cdot \Biggl[P(t) - P_0 \Biggr] \; = \; 
\frac{P^{\infty}_{ST}}{\tau_{ST}} \, \tau \,
\Biggl[1 \, - \, \exp{(-t/\tau)}\Biggr]
\label{eq:rise}
\end{equation}

\begin{figure}[hbt]
\centerline{\epsfig{file=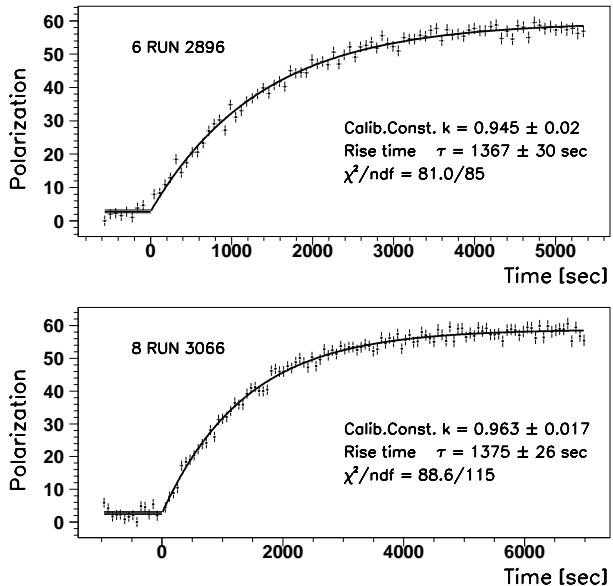, width=8cm, angle=0}}
\caption{Examples of rise-time calibration measurements}   
\end{figure}

The appropriate
input values for $P^{\infty}_{ST}$ and
$\tau_{ST}$ were listed in
table~1. The parameter $K$ is the calibration factor for the
polarization measurement of the polarimeter. The results of the 1996/97
rise-time fits are listed in table~2, together with the older calibration
data from 1994 which had been reported earlier~\cite{Barrow}.

\begin{table}[hbt]
\begin{center}
\begin{tabular}{cccccccc}\hline
Index & Year & Run & $K$ & $\Delta K$ & $\tau$ & $\Delta \tau$ &
$\chi^2/ndf$ \\
No.&      & No.  &       &       & (sec) & (sec) &   \\      
\hline
 1 & 1994 & 2370 & 0.936 & 0.021 &       &    &       \\
 2 & 1994 & 2442 & 0.976 & 0.042 &       &    &       \\
 3 & 1994 & 2444 & 0.900 & 0.038 &       &    &       \\
 7 & 1994 & 2482 & 0.960 & 0.030 &       &    &       \\
 8 & 1994 & 2484 & 0.956 & 0.023 &       &    &       \\
 9 & 1994 & 2486 & 0.955 & 0.028 &       &    &       \\
10 & 1994 & 2488 & 0.994 & 0.021 &       &    &       \\
11 & 1994 & 2492 & 0.902 & 0.026 &       &    &       \\ 
\hline
 1 & 1996 & 5138 & 0.971 & 0.042 & 1255  & 55 & 0.644 \\ 
 2 & 1996 & 7702 & 1.089 & 0.028 & 1637  & 42 & 0.788 \\
 3 & 1996 & 8382 & 1.006 & 0.023 & 1328  & 31 & 0.765 \\
 4 & 1997 & 2778 & 0.996 & 0.042 & 1392  & 61 & 0.680 \\
 5 & 1997 & 2824 & 0.962 & 0.032 & 1342  & 47 & 0.828 \\
 6 & 1997 & 2896 & 0.945 & 0.020 & 1367  & 30 & 0.953 \\
 7 & 1997 & 3030 & 0.998 & 0.031 & 1396  & 49 & 0.763 \\
 8 & 1997 & 3066 & 0.963 & 0.017 & 1374  & 25 & 0.770 \\
 9 & 1997 & 3278 & 0.906 & 0.052 & 1136  & 65 & 0.701 \\
10 & 1997 & 3316 & 0.999 & 0.043 & 1251  & 57 & 0.597 \\
11 & 1997 & 3318 & 1.030 & 0.070 & 1254  & 90 & 0.516 \\
14 & 1997 & 3669 & 0.933 & 0.039 &  873  & 37 & 0.821 \\
15 & 1997 & 6654 & 1.038 & 0.047 & 1172  & 53 & 0.922 \\
16 & 1997 & 6684 & 1.075 & 0.035 & 1362  & 49 & 0.784 \\
17 & 1997 & 6686 & 1.062 & 0.038 & 1388  & 53 & 0.838 \\
18 & 1997 & 6688 & 0.977 & 0.035 & 1265  & 49 & 0.702 \\
24 & 1997 & 9610 & 1.056 & 0.026 & 1391  & 37 & 0.795 \\
25 & 1997 & 9642 & 1.040 & 0.024 & 1463  & 36 & 0.507 \\
\hline
   &      &      &       &       &       &            \\
\end{tabular}
\caption{Calibration results from rise-time measurements}
\end{center}
\end{table}

In table 3 we give the weighted mean value of $K$, 
$\bar{K} = \sum w_i K_i /\sum w_i$ with $w_i = (\Delta K_i)^{-2}$,
the error of the mean value $\Delta \bar K = (\sum w_i)^{-1/2}$,
and the associated $\chi^2/ndf = \sum w_i \, (K_i - \bar K)^2 /(N-1)$.
Furthermore we give a rescaled error 
$\langle \Delta \bar K \rangle_{scaled} = 
(\chi^2/ndf)^{1/2} \cdot \Delta \bar K$
to account for the underestimation of the original errors indicated by 
$\chi^2/ndf > 1$. The scaling is equivalent to 
adding a common systematic error of size
$(\Delta \bar K)_{syst} = \Delta \bar K \cdot (\chi^2/ndf -1)^{1/2}$
in quadrature, see~\cite{pdg} for an explanation of these procedures.

\begin{table}[hbt]
\begin{center}
\begin{tabular}{ccccccc}\hline
data set & $\bar K_{rel}$  & $\bar K_{abs}$ &  $\Delta \bar K$  & 
$\chi^2/ndf$ & $(\Delta \bar K)_{syst} $ & $(\Delta \bar K)_{scaled} $  
\\
\hline
1994~    &  0.951  &  0.951  &  0.009  &  1.509  &  0.006 &  0.011  \\
1996/97  &  0.999  &  0.945  &  0.007  &  2.744  &  0.009 &  0.012  \\
\hline
         &         &         &        &         \\
\end{tabular}
\caption{Mean calibration factors and errors}
\end{center}
\end{table}

For a proper interpretation of the calibration results shown in table~3,
it is important to understand that all TPOL polarization values have been
scaled by a factor 0.946 since 1996~\cite{Barrow} in good agreement with
the value of 0.951 given here. The subsequent 1996/97 relative
re-calibration factor of 0.999 is therefore equivalent to an overall
absolute factor of $0.999 \cdot 0.946 = 0.945$ in relation to the original 
polarization scale used prior to 1996.

The analysis of the 1994 rise-time data reported in \cite{Barrow} assigned
an error of 0.032 to the determination of $K$. The error quoted is the rms
$\sigma$ of the nearly gaussian distribution of K-measurements. This rms
$\sigma$ value describes the typical error of a single measurement of $K$.
The error of the mean value of $K$ is then $\sigma /\sqrt{N}$ and thus
considerably smaller than $3\%$.

By comparing the calibrations of 1994 and 1996/97, we find excellent
agreement within the given errors of 1.1 and 1.2\%. Since the 
96/97-calibrations were carried out with activated spin rotators at
HERMES, in contrast to 1994 when the machine was flat, we can also set an
experimental upper limit of about 1.5\% on the correction $\Delta$
in equation~\ref{eq:pol-tau} which accounts for the
$\partial \hat{n}/ \partial \delta$ term in the numerator of
equation~\ref{eq:P}.

\section{LPOL/TPOL comparison}

As the magnitude of the polarization at any particular point in time
is an invariant around the HERA ring, a comparison between the TPOL
and LPOL polarimeters will provide a cross calibration of the instruments,
as long as the spin points fully upright at the TPOL and longitudinal at
the LPOL.
\begin{figure}[hbt]
\centerline{\epsfig{file=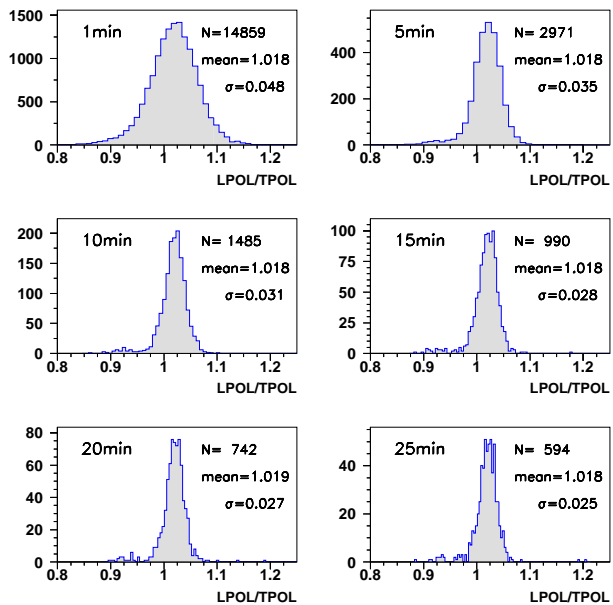, width=8cm, angle=0}}
\caption{LPOL/TPOL ratio measurements over a two-months period in 2000}
\end{figure}
Although it is well known to the operators of the polarimeters and to
the members of HERMES that the two instruments appeared to disagree
occasionally with each other outside of their quoted errors, it is 
reassuring to demonstrate a large sample of measurements, covering two
months of data taking in April/May of 2000, which exhibits a consistent
and stable performance of both polarimeters. We have plotted the LPOL/TPOL
ratio for this time period in figure~2 for six different integration times
ranging from 1 to 25 minutes. 
We obtain a consistent mean ratio of 1.018
with an error of the mean $\sigma /\sqrt N$ of less than 0.001.

Since  the statistical fluctuations are vanishing with higher 
averaging periods, one can estimate the systematic error of the TPOL
measurement $(\Delta P/P)_{TPOL}$ from the observed rms $\sigma = 0.025$
of the $25~min$ distribution and the quoted LPOL systematic error
$(\Delta P/P)_{LPOL} = 1.6\%$~\cite{lpol}. Assuming uncorrelated errors
for the two polarimeters, we obtain 
$(\Delta P/P)_{TPOL} = [\sigma^2 - (\Delta P/P)_{LPOL}^2]^{1/2} = 1.9\%$. 

\section{Acknowledgments}

The rise-time calibration analysis presented in this article 
and the LPOL/TPOL comparison is based on data collected by the
HERMES Polarimeter Group. 
We thank Desmond Barber for discussions on radiative polarization and
depolarization mechanisms.

\end{document}